\documentclass{article}
\usepackage{nips13submit_e,times}
\pdfoutput=1
\usepackage{url}
\usepackage{enumitem}
\usepackage[pdftex]{graphicx}
\usepackage[psamsfonts]{amssymb}
\usepackage{bm}                                     
\usepackage{amsmath,latexsym}               
\usepackage{caption}
\usepackage{subcaption}
\usepackage{microtype}

\usepackage[framemethod=latex]{mdframed}

\author{
Rasmus Troelsg\aa rd, Bj\o rn Sand Jensen and Lars Kai Hansen\\
Department of Applied Mathematics and Computer Science\\
Technical University of Denmark\\
Matematiktorvet 303B, 2800 Kgs. Lyngby\\
\texttt{\{rast,bjje,lkai\}@dtu.dk}\\
}

\nipsfinalcopy 

\title{A Topic Model Approach to Multi-Modal Similarity}

  \renewenvironment{thebibliography}[1]{%
    \begin{oldthebibliography}{#1}%
      \setlength{\parskip}{0ex}%
      \setlength{\itemsep}{0ex}%
  }%
  {%
    \end{oldthebibliography}%
}

\begin{document}
\maketitle

\begin{abstract} 
Calculating similarities between objects defined by many heterogeneous data modalities is an important challenge in many multimedia applications. We use a multi-modal topic model as a basis for defining such a similarity between objects. We propose to compare the resulting similarities from different model realizations using the non-parametric Mantel test. The approach is evaluated on a music dataset.
%
\end{abstract}
\section{Introduction}
%
%
Calculating similarity between objects linked to multiple data sources is more urgent than ever. A prime example is the typical multimedia application of music services where users face a virtually infinite pool of songs to choose from. Here choices are based on many different information sources including the audio/sound, meta-data like genre, and social influences \cite{salganik2006experimental}, hence, attempts of modeling the geometry of music navigation have taken on a multi-modal perspective. 
%
In fusing heterogeneous modalities like audio, genre, and user generated tags it is both a challenge to establish a combined model in a 'symmetric' manner so that one modality do not dominate others and it is challenging to evaluate the quality of the resulting geometric representation. Here, we focus on the latter issue by testing the consistency of derived inter-song (dis-)similarity by means of direct comparison between similarities using the Mantel permutation test.
%

Topic models have previously been used to infer geometry in the image and music domain, e.g. by \cite{Yoshii2006hybridplsa} combining audio features and listening histories. In \cite{Lienhart2009a} images and tags were analyzed, also by means of a multi-modal topic model. In \cite{Hoffman2008musichdp} music similarity is inferred with a nonparametric Bayesian model, and \cite{Blei2003annotated} describe multiple multi-modal extensions to basic LDA models and evaluate the models on an image information retrieval task. Furthermore, topic model induced similarities among documents have been put to use in a navigation application \cite{Chaney2003visualizingtm}, and different similarity estimates are also discussed in relation to a content-based image retrieval problem \cite{Horster2007}. 


\section{Model \& Inference}
\vspace{-0.1cm}
\begin{figure}[t]
\centering
\begin{subfigure}[b]{0.62\textwidth}  
\begin{mdframed}			
			\begin{small}
				\begin{itemize}[leftmargin=*]
				\item For each topic indexed by $t\in[1;T]$ in each modality indexed by $m\in[1;M]$\\
				Draw $\bm\phi_t^{(m)}\sim \textit{Dirichlet}(\bm\beta^{(m)})$\\
				This is the parameters of the $t^{th}$ topic's distribution over vocabulary $[1;V^{(m)}]$ of modality $m$.
				\item For each document indexed by $d\in[1;D]$
				\begin{itemize}
				\item Draw $\bm\theta_d \sim \textit{Dirichlet}(\bm\alpha)$ \\
				This is the parameters of the $d^{th}$ documents's distribution over topics $[1;T]$.
				\item For each modality $m\in[1;M]$
						\begin{itemize}
						\item For each word $w$ in the $m^{th}$ modality of document $d$
						\begin{itemize}
						\item Draw a specific topic $z^{(m)}\sim \textit{Categorical}(\bm\theta_d)$
						\item Draw a word $w^{(m)} \sim \textit{Categorical}(\bm\phi_{z^{(m)}}^{(m)})$
						\end{itemize}
						\end{itemize}
				\end{itemize}
			\end{itemize}
		\end{small}		
\end{mdframed}
\caption{Generative process}
\label{fig:minipage1}
\end{subfigure}
%
\quad
	\begin{subfigure}[b]{0.32\textwidth}
		\centering
			\includegraphics[width=1\textwidth, angle=90]{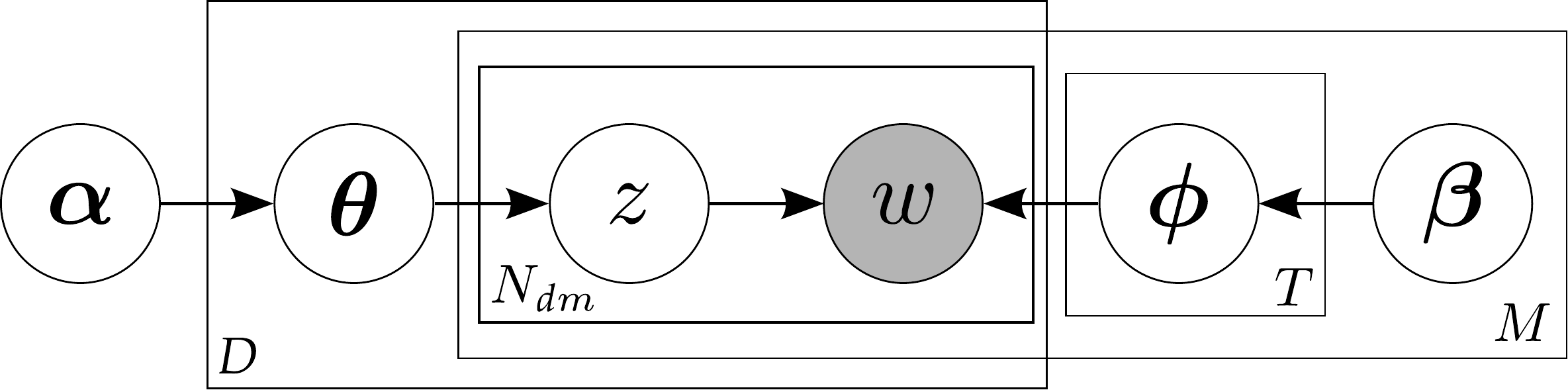}
			\caption{The multi-modal Latent Dirichlet Allocation model represented as a probabilistic graphical model.}
	\end{subfigure}
\caption{}
\label{fig:mmLDA_model}
	\vspace{-1.3em}
\end{figure}
%
%
%

To be able to measure similarities between objects, a representation of these objects is needed. In this work we use a version of Latent Dirichlet Allocation that incorporates multiple sources of information into a joint object representation similar to \cite{Blei2003annotated}. In \cite{polyling2009Mimno}, this model was applied to a multilingual corpus.\\
Each object is represented by a multinomial distribution over topics which is common for all of the modalities composing the object. Each topic is defined by a set of multinomial distributions over features, each of which is defined on the vocabulary specific for a modality. To explain the characteristics of the model, the assumed generative process for objects is outlined in figure \ref{fig:mmLDA_model} together with a graphical representation of the model. The difference from a number of individual LDA models, each defined on a separate modality, is that each object is described by a single, shared distribution over topics, which potentially induces strong dependencies between the feature distributions representing the same topic in the individual modalities.

Performing inference in the model amounts to estimation of the posterior distributions over the latent variables. We use a Gibbs sampler inspired by the sparsity improvements proposed by \cite{sparseLDA-Yao}. For evaluation (see section \ref{sec:result}), we use point estimates $\bm\theta^s$ and $\bm\phi^s$ derived from a sample $\mathbf z^s$ from the Markov chain, by taking the expectations of the respective posterior Dirichlet distributions defined by $\mathbf z^s$. In this work we choose the state of the chain with the highest model evidence within the last 50 out of 4000 iterations. Hyper-parameters are optimized using fixed point updates \cite{minka_polya,wallach_phd_thesis}. The prior on the document topic distributions is an asymmetric Dirichlet with parameter $\bm\alpha$, and the priors over the vocabularies of the respective modalities are symmetric Dirichlet distributions with parameters $\bm\beta^{(m)}$.
\section{Similarities in Topic Models}
\vspace{-0.1cm}
As already hinted, there are many ways to define and calculate similarities in topic models; both between topics and documents. In this paper we focus on the latter. Most methods in literature are based solely on the distributions of topics in the documents, $\bm\theta$, e.g. \cite{Hoffman2008musichdp} measures the Kullback-Leibler divergence between two such distributions, while \cite{Horster2007} also mentions inner products and cosine similarities as candidates. With focus on visualization, \cite{Chaney2003visualizingtm}, introduces the yet another dissimilarity measure based on topic proportions. \cite{Horster2007} promotes a measure based on the predictive likelihood of the document contents, and this approach is the basis of the method chosen here;
The similarity of two documents $A$ and $B$ is given by the mean per-word log-likelihood of the words of document $A$ given the topic distribution of document $B$ (and the vocabulary distributions).
\begin{align}
%
\frac{\log p(\mathbf w_A|\bm\theta_B^s,\bm\phi^s)}{\sum_{m=1}^MN^{(m)}_A}, \text{\,\,\,\,where\,\,\,\,} p(\mathbf w_A|\bm\theta_B^s,\bm\phi^s)=\prod_{m=1}^M\prod_{i=1}^{N_A^{(m)}}\sum_{t=1}^T{(\bm\phi^{(m)}_{t,w^{(m)}_{Ai}})}^{\top}\bm\theta_{t,B}
\end{align}
We use this approach to calculate a non-symmetric similarity matrix between all objects in the held-out cross-validation fold, for which the topic proportions have been estimated using ``fold-in''.
\footnote{For the few held-out documents that do not contain any words in the modalities used for model estimation, we chose to simulate a uniform distribution of words in such an empty document by one occurrence of every word in the vocabulary.}
While this similarity measure is more computationally demanding than e.g. the KL-divergence, when the number of topics $T$ used in the model increases, it might happen that some topics have vocabulary distributions that are very alike and only differ on a few words. Thus two documents with mainly the same type of content may have large proportions of different topics, causing them to be very dissimilar according to a topic proportion based measure. For a non-parametric topic model such as  \cite{Hoffman2008musichdp}, this might not be a large concern, however, for parametric topic models, this should be taken into consideration. Generally, most of the discussed similarity measures are not proper metrics in the geometric sense, but for (dis-)similarity purposes the exact properties might not be important, depending on the application.
%

\textbf{Comparing Similarities - the Mantel test}\\
An important aspect of this work is the ability to assess the relations between different similarities induced by models estimated from multiple, possibly different, heterogeneous data sources. To compare such similarities we look at the correlation between the defined similarities. For testing the significance of the correlations we can apply a Mantel style test \cite{Mantel1967test}. The Mantel test is a non-parametric test to assess the relation between two (dis-)similarity matrices. 
The null hypothesis is that the two matrices are unrelated, and the null distribution is approximated by calculating the test statistic for a large number of random permutations of the two matrices (excluding the diagonal elements); permuting rows and columns together to maintain the distribution of (dis-)similarities for each object. In this work we use Spearman's correlation coefficient as the test statistic.

\section{Experimental Results: Music Similarity}\label{sec:result}
\begin{figure}[b]
  \centering
  \begin{subfigure}[b]{0.41\textwidth}
    \includegraphics[width=1\textwidth]{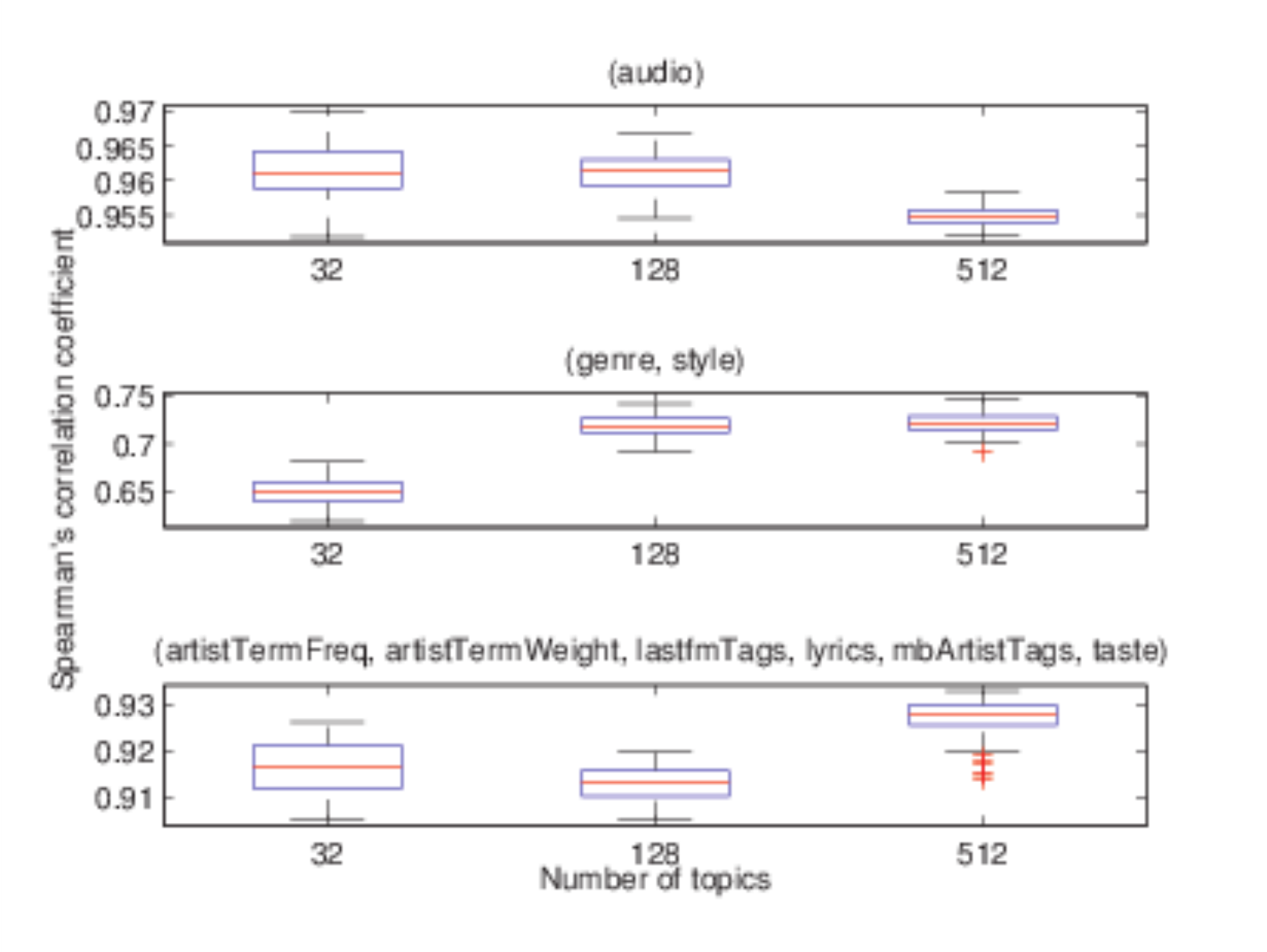}
  \caption{{\small Boxplots of Spearman's correlation between similarity matrices obtained using the same parameters and data, but different random initialisation. Each row correspond to a modality group.}}
  \label{fig:self_corr}
  \end{subfigure}
\qquad
  \begin{subfigure}[b]{0.41\textwidth}
    \includegraphics[width=1\textwidth]{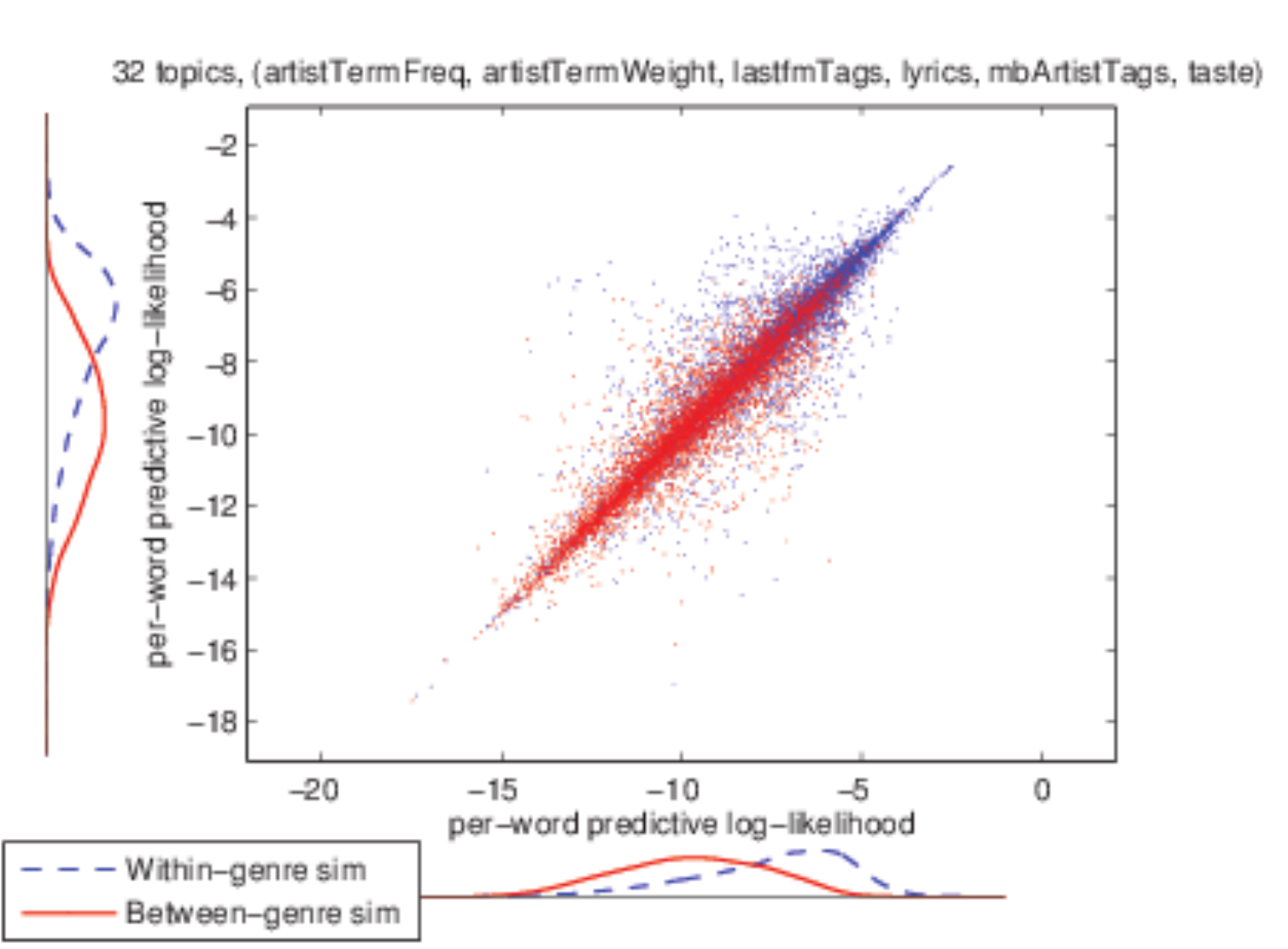}
    \caption{{\small Scatter plot of two specific similarities obtained by random initialisations of the same model estimated from the larger group of modalities. The colors indicate within- and between-genre similarities.}}
  \end{subfigure}
\caption{}
\label{fig:stability}
\end{figure}

\begin{figure}[t]
  \centering
  \begin{subfigure}[b]{0.41\textwidth}
    \includegraphics[width=1\textwidth]{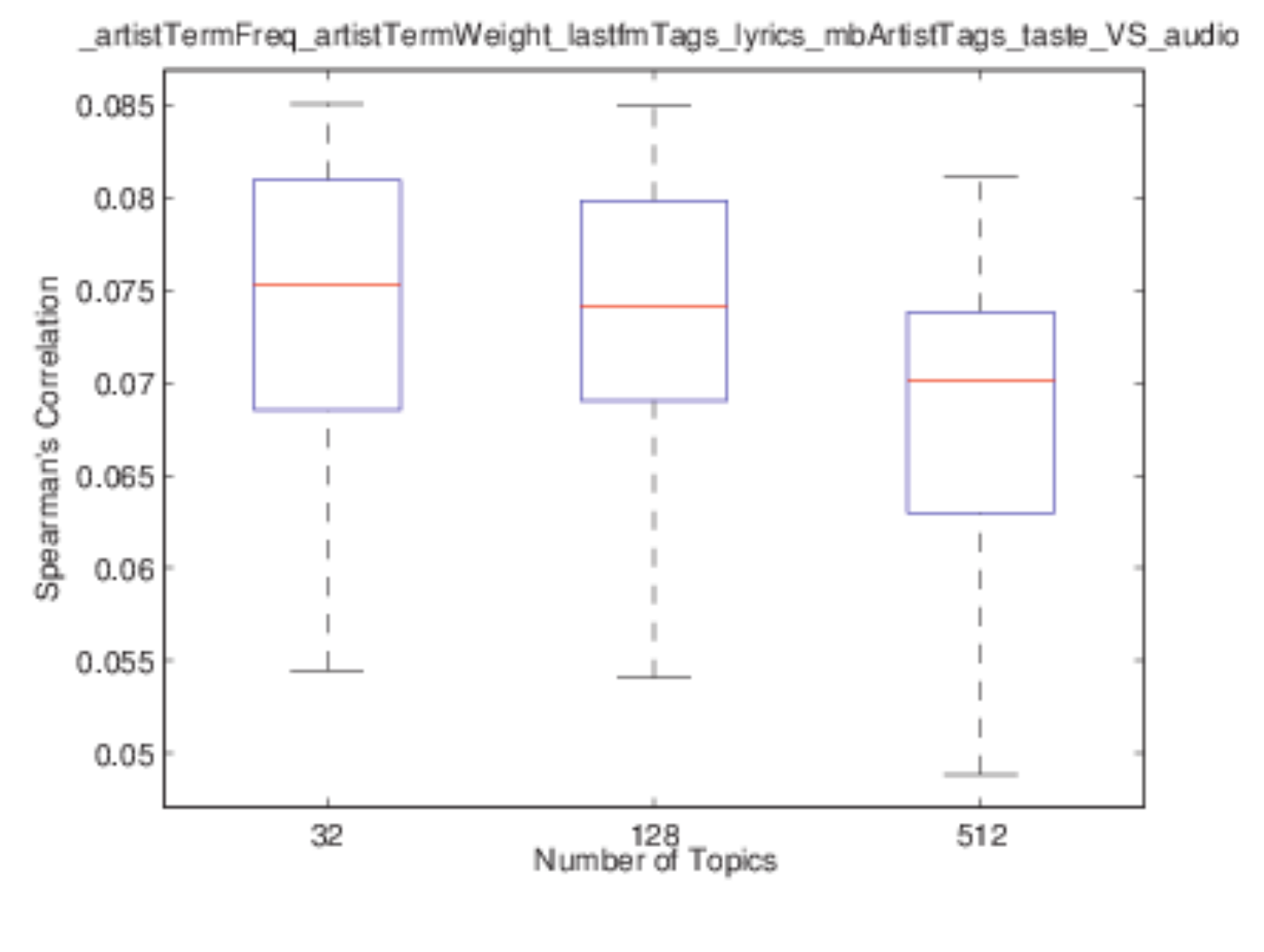}
    \caption{{\small Boxplots of Spearman's correlation between similarity matrices obtained using the larger modality group and the audio.}}
\label{fig:meta_cross}
  \end{subfigure}
  \qquad
  \begin{subfigure}[b]{0.41\textwidth}
    \includegraphics[width=1\textwidth]{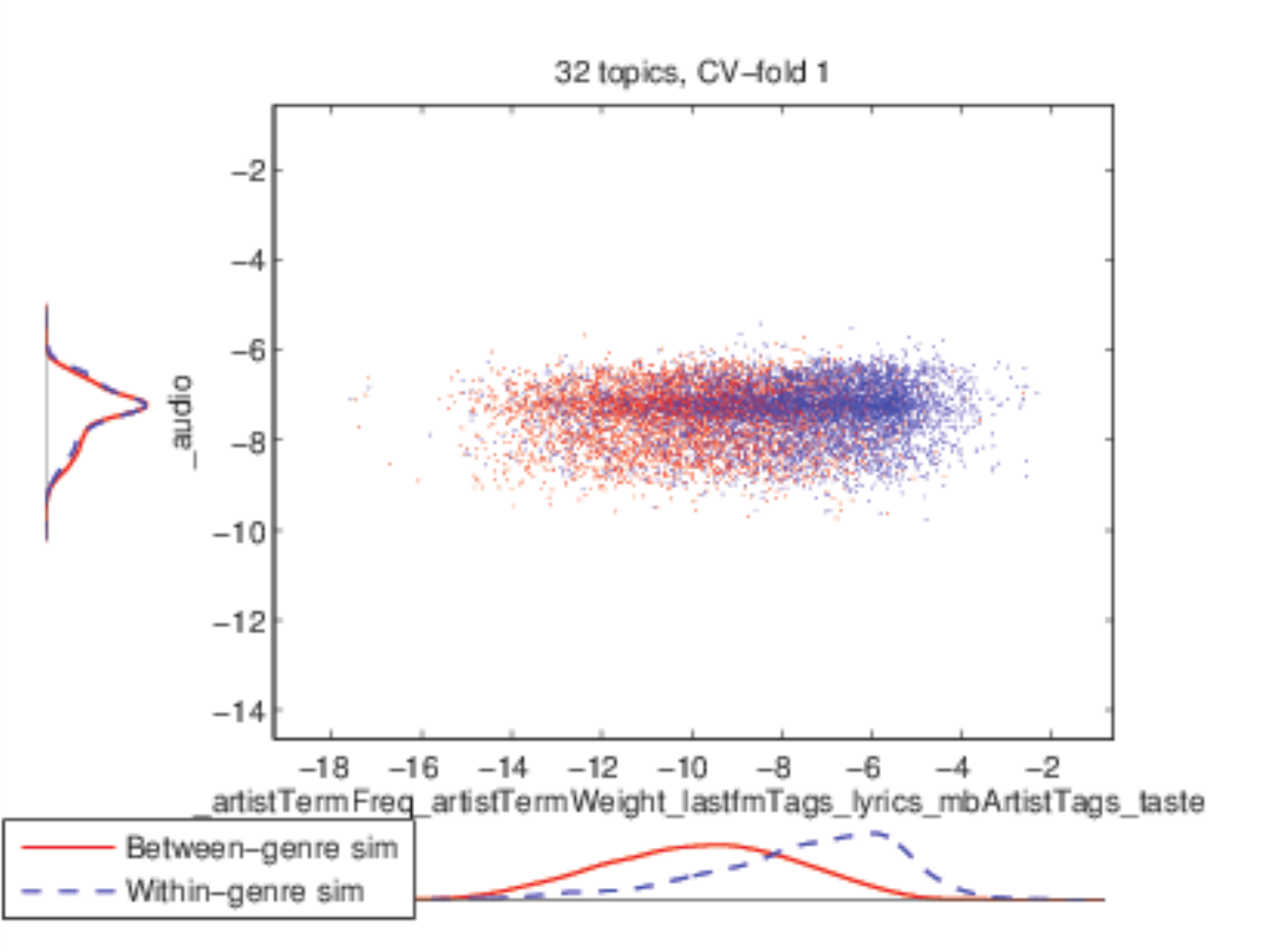}
    \caption{{\small Scatter plot of the two examples of similarities obtained from topic models with same parameters, but two mutually exclusive modalities of data.}}
    \label{fig:meta_cross_scatter}
  \end{subfigure}
\caption{}
\label{fig:cross}
\end{figure}

In this preliminary study we examine induced similarities in a subset of the Million Song Dataset~\cite{Bertin-Mahieux2011}, consisting of 30.000 tracks with equal proportions of 15 different genres. Each track is composed of data from a number of different sources:
Open vocabulary tags from users (last.fm),
Lyrics (musiXmatch.com),
Editorial artist tags (allmusic.com),
Artist tags (musicBrainz),
User listening history (echonest),
Genre and style (allmusic), and 
Audio Features (echonest).
All modalities---besides the audio features---are naturally occurring as counts of words and for the audio we turn to an \textit{audio word} approach, where the continuous features are vector quantized into a total of 2144 words.
For this pilot study we estimate topic models on combinations of groups of modalities from the mentioned list, respectively consisting of the first 5, the genre and style labels, and the audio. To be able to assess the model stability of the similarities, we estimate each model five times from different random initialisations of the Markov chain. This is done for every training set of a 10-fold cross-validation split. The correlations between all combinations of the 5 similarity matrices resulting from each held-out fold are then calculated, and the resulting distributions of correlation coefficients are shown in figure \ref{fig:self_corr}. Figure \ref{fig:meta_cross} shows the distributions of correlations between similarities based on audio and on the larger modality group. The correlations are evidently much smaller than for identical models, but a Mantel test with 100 permutations suggest that the null hypothesis of no correlation can be rejected at a significance level of at least 1\% for all three model complexities.
%
%

\section{Discussion \& Conclusion}
%
%
%
%
%
The issue of stability is relevant for similarities induced by topic models using approximate inference techniques. The correlations between similarities from identical but randomly initialized models, can be used as a tool to gain some insight into this matter. From the preliminary results on the music example we find the induced similarities (fig. \ref{fig:stability}) to be highly stable.
Furthermore, inspecting the similarities obtained from different data types; figure \ref{fig:cross}, we observe that while the audio model in itself does not seem to provide higher intra- than inter-genre similarity, it is still significantly positively correlated to the other modality group which does possess some discriminative power in terms of genre labels. Moreover, it seems that an increasing number of topics causes the correlation between similarities from models estimated on different modality groups to decrease. We speculate that this is linked to the specific topic model variant, for which \cite{Blei2003annotated} also note that the model describes the joint distribution of different modalities well, but does not model the relations between them.

In conclusion, we have proposed the multi-modal LDA as a method to define similarities in \hyphenation{multi-media}multimedia applications with multiple heterogeneous data sources based on the predictive-likelihood. This was extended with the Mantel test allowing direct evaluation of the consistency and correspondence of the resulting similarities.


\subsubsection*{Acknowledgment} 
\vspace{-3.5mm}
This work was supported in part by the Danish Council for Strategic Research of the Danish Agency for Science Technology and Innovation under the CoSound project, case number 11-115328. This publication only reflects the authors' views.

\bibliographystyle{unsrt}
\renewcommand\refname{\vskip -0.7cm}
\subsubsection*{References}
\small
\bibliography{refs}

\end{document}